\documentclass[prl,twocolumn,showpacs,superscriptaddress,nofootinbib,floatfix]{revtex4-2}

\usepackage[utf8]{inputenc}
\usepackage{amsmath,amssymb}
\usepackage{slashed}
\usepackage{subfigure}
\usepackage[colorlinks=true, 
linkcolor=blue,
breaklinks=true,
urlcolor=magenta,
citecolor=blue]{hyperref}
\usepackage[usenames,dvipsnames]{color}
\usepackage{appendix}
\usepackage{braket,bm}
\usepackage{multirow}
\usepackage{enumitem}
\usepackage{color}
\usepackage{cancel}
\usepackage{orcidlink}
\usepackage{mathrsfs}
\usepackage{graphicx}
\usepackage[normalem]{ulem}
\usepackage{booktabs}
\usepackage{array}
\usepackage{overpic}
\usepackage{float}
\usepackage{threeparttable}
\usepackage{dsfont}
\usepackage{soul}
\allowdisplaybreaks[4]

\newcommand{\itp}{\affiliation{Institute of Theoretical Physics, Chinese Academy of Sciences, Beijing 100190, China}}

\newcommand{\ucas}{\affiliation{School of Physical Sciences, University of Chinese Academy of Sciences, Beijing 100049, China}}

\newcommand{\scnt}{\affiliation{Southern Center for Nuclear-Science Theory (SCNT), Institute of Modern Physics,\\ 
Chinese Academy of Sciences, Huizhou 516000, China}}

\newcommand{\huu}{\affiliation{School for Theoretical Physics, School of Physics and Electronics, Hunan University, Changsha 410082, China}}

\newcommand{\huhep}{\affiliation{Hunan Provincial Key Laboratory of High-Energy Scale Physics and Applications,\\
Hunan University, Changsha 410082, China}}

\newcommand{\md}{\mathrm{d}}

\graphicspath{{fig/}}

\begin{document}
\title{Precision Sum Rule for Nucleon Isovector Polarizabilities and\\ the Proton--Neutron Mass Difference} 

\author{Xiong-Hui Cao\orcidlink{0000-0003-1365-7178}}\email{xhcao@itp.ac.cn}
\itp

\author{Ling-Yun Dai\orcidlink{0000-0002-4070-4729}}
\email{dailingyun@hnu.edu.cn}
\huu\huhep

\author{Feng-Kun Guo\orcidlink{0000-0002-2919-2064}}\email{fkguo@itp.ac.cn}
\itp\ucas\scnt

\begin{abstract}

The precision of the electromagnetic proton--neutron mass difference $\delta m_\mathrm{QED}$ extracted from the Cottingham formula hinges on a subtraction function whose low-energy normalization $\bar S(0)$ is fixed by the isovector combination of the proton and neutron  polarizabilities, $(\alpha_{E1}-\beta_{M1})^{p-n}$.
We derive a dispersive sum rule for this combination in terms of $s$-channel photoabsorption cross sections and the product of $t$-channel $\gamma\gamma\to\pi\eta/K\bar K_{I_t=1}$ and $\pi\eta/K\bar K_{I_t=1}\to N\bar N$ amplitudes, without invoking Reggeon dominance.
Combining empirical pion-photoproduction multipoles with coupled-channel Muskhelishvili--Omn\`es representations of the scalar-isovector $\pi\eta/K\bar K$ amplitudes, we obtain $(\alpha_{E1}-\beta_{M1})^{p-n}=-0.82(33)\times10^{-4}\,\mathrm{fm}^3$, fixing its sign and reducing the uncertainty by one order of magnitude compared with the previously known value.
This result yields $\bar S(0)=-0.56(28)\,\mathrm{GeV}^{-2}$, leading to $\delta m_\mathrm{QED}=0.75^{+0.02}_{-0.04}\,\mathrm{MeV}$, substantially more precise than previous Cottingham determinations.
The negative $\bar S(0)$ also provides a low-energy test of Reggeon dominance in the subtraction function.
\end{abstract}

\maketitle

\section{Introduction}

The electric and magnetic dipole polarizabilities of the nucleons, $\alpha_{E1}$ and $\beta_{M1}$, are fundamental structure constants that quantify their resistance to deformation by quasi-static electromagnetic fields. 
As the leading non-pointlike electromagnetic response of a composite particle, they provide a sensitive probe of the dynamics that govern the nucleon internal structure.
Recent advances in real Compton scattering (RCS) experiments have enabled the first simultaneous extraction of all six leading-order proton polarizabilities~\cite{Mornacchi:2022cln}. 
These quantities are not only of intrinsic importance for nonperturbative quantum chromodynamics (QCD), particularly as an important test of chiral dynamics~\cite{Bernard:1991rq,Bernard:1993bg,Lvov:1993ex}, but also play a vital role in precision atomic physics and astrophysical applications, e.g., contribution to two-photon-exchange effects in the Lamb shift and hyperfine structure~\cite{Pachucki:1996zza,Pachucki:1999zza,Carlson:2011zd,Hill:2011wy,Birse:2012eb}, as well as constraints relevant to neutron-star properties~\cite{Bernabeu:1974zu}; see Refs.~\cite{Bernard:1995dp,Drechsel:2002ar,Schumacher:2005an,Griesshammer:2012we,Holstein:2013kia,Hagelstein:2015egb} for reviews.

While the proton polarizabilities are now being mapped with increasing accuracy, the neutron ones continue to pose a challenge due to the lack of free neutron targets. The current  Particle Data Group (PDG) averages are~\cite{ParticleDataGroup:2026mpi}
\begin{align}
    \begin{aligned}
        \alpha_{E1}^{p}&=11.5(4),\quad &\beta_{M1}^{p}&=2.31(29), \\ 
        \alpha_{E1}^{n}&=11.8(1.1), & \beta_{M1}^{n}&=3.7(1.2),
    \end{aligned}
    \label{eq:polvalues}
\end{align}
in units of $10^{-4}\,\mathrm{fm}^3$ hereafter.
The isovector differences, $\alpha_{E1}^{p-n}$ and $\beta_{M1}^{p-n}$, are numerically small as they vanish in the isospin-symmetric limit.
Yet, it is precisely this suppression that renders them an exceptionally clean probe.
The isovector electric polarizability $\alpha_{E1}^{p-n}$ provides a sensitive test of the Reggeon-dominance hypothesis for the high-energy behavior of Compton amplitudes~\cite{Gasser:2015dwa,Capella:1994cr,Kaidalov:1998pn,Alwall:2004wk}.
Furthermore, poor knowledge of $\beta_{M1}^{p-n}$ has dominated the uncertainty in the electromagnetic contribution to the proton--neutron mass difference~\cite{Walker-Loud:2012ift,Thomas:2014dxa,Erben:2014hza,Gasser:2015dwa,Tomalak:2018dho}. 

The proton--neutron mass difference is a fundamental quantity. It is a characteristic isospin-breaking observable and is precisely measured~\cite{Mohr:2024kco}, 
\begin{align}
    \label{eq:CODATA} m_p-m_n=-1.29333251(38)~\mathrm{MeV}, 
\end{align} 
which is essential for precision treatments of primordial nucleosynthesis, where this mass difference enters the $\beta$-decay rates~\cite{heffernan2017quantifying,Meissner:2023voo,Meyer:2024auq}.
It reflects a fine-tuned balance between the QCD light-quark mass difference $m_d-m_u$ and electromagnetic effects, on which hinges the existence of stable nucleons, nuclei, atoms, and stars~\cite{Hogan:1999wh}. 
Decomposing it into the strong contribution $\delta m_\mathrm{QCD}$ and the electromagnetic contribution from quantum electrodynamics (QED) $\delta m_\mathrm{QED}$, which have comparable magnitudes and opposite signs, is a long-standing problem. This decomposition is essential for determining the isovector (pseudo)scalar charge $g_{S(P)}^{p-n}$~\cite{Gonzalez-Alonso:2013ura}.

The leading electromagnetic correction $\delta m_\mathrm{QED}$ can be given by the Cottingham formula~\cite{Feynman:1954xxl,Chou1959mass,Cini:1959szx,Cottingham:1963zz}, which relates it to an integral over the forward doubly-virtual Compton scattering (VVCS) amplitude of the nucleon.\footnote{Chou~\cite{Chou1959mass} and Cini et al.~\cite{Cini:1959szx} independently used dispersion relations to relate the electromagnetic mass difference to the VVCS amplitude. 
Both works concluded that the dominant elastic (Born) term alone cannot reproduce the observed mass difference.}
Practical evaluations of this formula, however, are limited by the need for a subtraction function that is not directly accessible in experiment~\cite{Harari:1966mu}.
The uncertainty in the subtraction function has been dominated by the poor knowledge of $\beta_{M1}^{p-n}$, as recently pointed out~\cite{Walker-Loud:2012ift,Thomas:2014dxa,Erben:2014hza,Leutwyler:2015jga,Tomalak:2018dho,Walker-Loud:2019qhh}. 
Traditional extractions of $\beta_{M1}^{p-n}$ from Compton cross-section data have yielded large uncertainties (often $\gtrsim 100 \%$), limiting the precision of $\delta m_{\mathrm{QED}}$.
There have also been lattice results on the decomposition of the proton--neutron mass difference~\cite{Blum:2010ym,BMW:2014pzb,Endres:2015gda,CSSM:2019jmq}; however, the Cottingham analysis and lattice calculations employ different schemes for separating QED and QCD effects; see Ref.~\cite{Gasser:2003hk} and Sec.~4.15 of Ref.~\cite{Meissner:2022cbi}.

In this work, we link $(\alpha_{E1}-\beta_{M1})^{p-n}$ directly to low-energy data, which are the $s$-channel photoabsorption cross sections and $t$-channel $\gamma\gamma\to\pi\eta/K\bar K_{I_t=1}$ and $\pi\eta/K\bar K_{I_t=1}\to N\bar N$ amplitudes, by a new sum rule based on dispersion relations. 
The construction avoids the assumption of high-energy Reggeon dominance~\cite{Elitzur:1970yy,Gasser:1974wd,Gasser:2015dwa}.
Consequently, we are able to determine $(\alpha_{E1}-\beta_{M1})^{p-n}$ with a substantially improved precision. 
This result supplies the low-energy input for a Cottingham-formula evaluation, leading to an improved dispersive determination of $\delta m_\mathrm{QED}$.

\section{Isovector polarizability sum rules for the nucleon}

The electric and magnetic dipole polarizabilities, $\alpha_{E1}$ and $\beta_{M1}$, can be accessed experimentally through low-energy RCS. 
At leading order in the expansion of the incident photon energy $\omega$ in the laboratory frame, the response is described by the second-order effective Hamiltonian
\begin{align}
    H_{\text{eff}}^{(2)}= -4 \pi\left(\frac{1}{2} \alpha_{E1} \mathbf{E}^2+\frac{1}{2} \beta_{M1} \mathbf{H}^2\right).
\end{align}
A well-established constraint is provided by the Baldin sum rule~\cite{baldin1960polarizability,lapidus1963scattering}, $\alpha_{E1}+\beta_{M1}=\frac{1}{2 \pi^2} \int_{0}^{\infty} \md\omega~\sigma_{\text{tot}}(\omega)/\omega^2$, which relates the sum to the total photoabsorption cross section. 
Thanks to significant experimental progress, particularly from modern polarized photon beam facilities~\cite{A2:2021wmm,Li:2022vnz}, precise determinations of the proton polarizabilities have become possible.
However, neutron polarizabilities must be inferred from nuclear targets, and nuclear effects are involved. 
Consequently, the isovector polarizabilities are poorly constrained: the PDG averages~\cite{ParticleDataGroup:2026mpi} in Eq.~\eqref{eq:polvalues}, with the proton and neutron uncertainties added in quadrature, give $\alpha_{E1}^{p-n}=-0.3(1.2)$ and $\beta_{M1}^{p-n}=-1.4(1.2)$, leaving even their signs uncertain.

The large uncertainties motivate a more direct approach. 
Dispersion relations (DRs) for RCS provide a model-independent link between polarizabilities and measurable amplitudes. 
In standard subtracted fixed-$t$ DRs, however, the dipole polarizabilities reside in subtraction constants rather than being predicted~\cite{Drechsel:1999rf}. 
They must therefore be fitted to RCS data, inheriting the low-energy experimental uncertainties; even the definition of the proton RCS data set below pion-production threshold remains a matter of ongoing debate~\cite{Krupina:2017pgr,Pasquini:2019nnx}.

We use the fixed-$su$ or hyperbolic DR~\cite{Hearn:1962zz,Hite:1973pm,Hohler:1983}, with $s$, $u$ and also $t$ the Mandelstam variables. 
This framework incorporates crossing symmetry and analyticity and treats the discontinuities from both the $s$-($u$-)channel process $\gamma N \rightarrow \gamma N$ and the annihilation channel $\gamma \gamma \rightarrow \bar{N} N$ consistently. 
The latter encodes the isoscalar and isovector meson-exchange physics essential for the polarizability difference.
Considering the proton--neutron difference $(\dots)^{p-n} \equiv (\dots)^p - (\dots)^n$, the dispersive framework leads to an isovector sum rule,
\begin{align}\label{eq:SR_isovector}
    \left(\alpha_{E1}-\beta_{M1}\right)^{p-n}=(\alpha_{E1}-\beta_{M1})^s+2(\alpha_{E1}-\beta_{M1})^t\,,
\end{align}
where $(\alpha_{E1}-\beta_{M1})^s$ is the total $s$-channel contribution to the difference, and $(\alpha_{E1}-\beta_{M1})^t$ denotes the isovector $t$-channel term. 
The factor of $2$ reflects the isovector nature of the $t$-channel exchange, where the proton and neutron enter with opposite signs.
Here, the $s$-channel part reads~\cite{Bernabeu:1974zu}
\begin{align}\label{eq:SR_isovector_s}
    (\alpha_{E1}-\beta_{M1})^{s}=~&\frac{1}{2 \pi^2} \int_{M_\pi+\frac{M_\pi^2}{2 m_N}}^{\infty}\frac{\mathrm{d} \omega}{\omega^2}~\sqrt{1+\frac{2 \omega}{m_N}}\nonumber\\
    &\times[\sigma(\Delta P=\mathrm{yes})-\sigma(\Delta P=\mathrm{no})]\,,
\end{align}
where photoexcitation cross sections that preserve the nucleon parity ($\Delta P = \mathrm{no}$) enter with an opposite sign relative to those that flip it ($\Delta P = \mathrm{yes}$), and each $\sigma$ should also be understood as for $p-n$.
The $t$-channel contribution $(\alpha_{E1}-\beta_{M1})^t$ is nearly saturated by the isovector two-meson continua $\gamma\gamma\to \pi\eta/K \bar K_{I_t=1}\to N\bar N$ (see End Matter for details),\footnote{We neglect the $D$-wave, dominated by the $a_2(1320)$ resonance.
For the minimal-coupling term $T_{\mu\nu} F^{\mu\rho} F^{\nu}_{\ \rho}$, with $T_{\mu\nu}$ a symmetric traceless tensor for the $a_2(1320)$ and $F^{\mu\rho}$ the electromagnetic field tensor, the equal-helicity photon amplitudes vanish~\cite{Drechsel:1999rf}.
This is supported empirically by Belle analyses, which found negligible $a_2(1320)$ contribution to equal-helicity amplitudes~\cite{Belle:2009xpa,Belle:2013eck}.}
\begin{align}\label{eq:SR_isovector_t}
    &(\alpha_{E1}-\beta_{M1})^t_{\pi\eta/K \bar K_{I_t=1}}\nonumber\\
    ={}&\frac{8\alpha_\mathrm{em}}{\pi} \int_{t_{\pi\eta}}^{\infty}\frac{\mathrm{d} t}{t^2}\frac{1}{4 m_N^2-t}\sigma_{\pi\eta}(t) g_+^{0}(t)F^{*}_{\pi\eta,0\,0}(t) \nonumber\\
    &\frac{8\sqrt{2}\alpha_\mathrm{em}}{\pi} \int_{t_K}^{\infty}\frac{\mathrm{d} t}{t^2}\frac{1}{4 m_N^2-t}\sigma_{K\bar K}(t) h_+^{0,1}(t)F^{1\,*}_{K,0\,0}(t)\,,
\end{align}
where $I_t$ is the $t$-channel isospin, $t_{\pi\eta}=(M_\pi+M_\eta)^2$, $t_K=4M_K^2$, $\alpha_\mathrm{em}$ is the fine-structure constant, and $m_N,M_\pi,M_\eta$ denote the nucleon, pion, and eta masses, respectively. 
The phase-space factor is $\sigma_{A B}(t)=\lambda^{1/2}(t,M_A^2,M_B^2)/t$, with $\lambda(a, b, c)=(a-b-c)^2-4 b c$ the K\"all\'en function.
Here, $g_{+}^{J=0}$ and $h_{+}^{J=0,I=1}$ are the $S$-wave isovector amplitudes for $\pi\eta \to N\bar{N}$ and $K\bar{K} \to N\bar{N}$, respectively. 
The subscript $+/-$ refers to parallel/antiparallel antinucleon-nucleon helicities.
Similarly, $F_{\pi\eta,J=0,\Lambda_\gamma=0}$ and $F_{K,J=0,\Lambda_\gamma=0}^{I=1}$ represent the $S$-wave isovector amplitudes for $\gamma\gamma\to \pi\eta$ and $\gamma\gamma\to K\bar{K}$.
The subscript $\Lambda_\gamma=0$ denotes parallel two-photon helicities.
Equations~\eqref{eq:SR_isovector_s} and~\eqref{eq:SR_isovector_t} are the isovector counterparts of the Bernab\'eu--Ericson--Ferro Fontan--Tarrach sum rule~\cite{Bernabeu:1974zu,Bernabeu:1977hp}.

\section{Muskhelishvili--Omn\`es representation}

Unlike the Baldin sum rule, the $s$-channel contribution in Eq.~\eqref{eq:SR_isovector_s} cannot be obtained directly from the total photoabsorption cross section, because of the required parity separation.
The dominant contribution comes from the well-established single-pion channel, $\gamma N\to \pi N$.
We evaluate it using multipole amplitudes from three comprehensive partial-wave analyses: SAID-2023~\cite{Briscoe:2023gmb,GWU}, BnGa-2019~\cite{Anisovich:2016vzt,BG}, and MAID-2007~\cite{Drechsel:2007if,Mainz}. 
Taking the average as the central value gives $(\alpha_{E1}-\beta_{M1})^s_{\pi N}=-1.27(25)\times10^{-4}$. 
The uncertainty is estimated conservatively from the largest difference among the three analyses.
\footnote{As an independent cross-check, evaluating this contribution with the ANL-Osaka-2019 dynamical coupled-channel (DCC) multipoles~\cite{Kamano:2019gtm,ANL-O} yields $-1.43$, fully consistent with our estimate within uncertainties; the J\"uBo-2016 DCC model~\cite{Anisovich:2016vzt,JuBo}, for which neutron multipoles are unavailable, likewise yields a proton contribution compatible with our proton-target evaluation.}

The remaining $s$-channel contribution from non-$\pi N$ inelastic final states, primarily $\pi\pi N$, is decomposed into a nonresonant background and inelastic resonant contributions~\cite{Lvov:1996rmi}. 
The nonresonant background is dominated by the $\pi\Delta$ intermediate state, which, within the one-pion-exchange approximation, acts identically on the proton and neutron due to isospin symmetry~\cite{Holstein:1994tw,Drechsel:1999rf}, thereby dropping out of Eq.~\eqref{eq:SR_isovector}. 
Inelastic resonant contributions, such as those from the $N^*(1535)$, are estimated by assuming the same multipole structure as in single-pion photoproduction with appropriate rescaling~\cite{Lvov:1996rmi}.
Using the corresponding resonant photo-decay parameters~\cite{Workman:2011vb,Workman:2012jf}, we find $(\alpha_{E1}-\beta_{M1})^s_\text{inel}=0.07\times 10^{-4}$; even with an estimated uncertainty of $\lesssim 100\%$, it remains an order of magnitude below the single-pion uncertainty. 
We thus include it without attaching an extra error.

The $t$-channel contribution requires partial-wave amplitudes for $\gamma \gamma \rightarrow \pi \eta/K \bar{K}_{I_t=1}$ and $\pi \eta/K \bar{K}_{I_t=1} \rightarrow N \bar{N}$. 
The main challenge is the lightest scalar-isovector state, the $a_0(980)$ (see, e.g., Refs.~\cite{Baru:2003qq,Dai:2011bs}).
Because the $a_0(980)$ couples strongly to $K \bar{K}$, a coupled-channel dispersive analysis of $\pi \eta$ and $K \bar{K}$ is essential.
This strategy has been successfully applied in the isoscalar channel, $\gamma\gamma \to \pi\pi/K\bar{K}_{I_t=0}$~\cite{Pennington:2006dg,Oller:2008kf,Mao:2009cc,Garcia-Martin:2010kyn,Hoferichter:2011wk,Dai:2014zta,Dai:2016ytz,Danilkin:2018qfn}.
The relevant two-photon amplitudes satisfy a modified Muskhelishvili--Omn\`es (MO) representation~\cite{Lu:2020qeo},
\begin{align}\label{eq:MO_rep}
    \mathbf{F}_{0\,0}^1(t)={}&\mathbf{F}_{0\,0}^{1,\rm{Born}}(t)+t\mathbf{\Omega}^1_0(t)\Biggl[\mathbf{P}_0(t)\nonumber\\
    & +\frac{t-t_0}{\pi}\int_{\mathcal{C}_L}\md t^\prime~\frac{\left[\mathbf{\Omega}^1_0(t')\right]^{-1} \operatorname{Im}\mathbf{F}_{0\,0}^{1,\rm{Res}}(t')}{t'(t'-t_0)(t'-t)}\nonumber\\
    & -\frac{t-t_0}{\pi}\int_{t_{\pi\eta}}^{\infty}\md t^\prime~\frac{ \operatorname{Im}\left[\mathbf{\Omega}^1_0(t')\right]^{-1}\mathbf{F}_{0\,0}^{1,\rm{Born}}(t')}{t'(t'-t_0)(t'-t)}\Biggr],
\end{align}
where $\mathbf{F}_{0\,0}^1\equiv \left(F_{\pi\eta,0\,0}, F^1_{K,0\,0}\right)^T$, $\mathbf{P}_0(t)$ is a vector-valued subtraction polynomial fixed to satisfy Low's theorem in the soft-photon limit~\cite{Low:1954kd}, $\mathcal{C}_L$ denotes the integration path along the resonance-exchange left-hand cuts, and $\mathbf{F}_{0\,0}^{1,\rm{Res}}$ contains the left-hand-cut contributions excluding the Born term,
\begin{align}
    \mathbf{F}_{0\,0}^{1,\rm{Born}}(t)=
    \begin{pmatrix}
        0  \\
        -\frac{\sqrt{2} M_K^2}{t} \frac{1}{\sigma_{K\bar K}(t)}\log\frac{1+\sigma_{K\bar K}(t)}{1-\sigma_{K\bar K}(t)} 
    \end{pmatrix}.
\end{align}
Through the isovector $S$-wave Omn\`es matrix $\mathbf{\Omega}^1_0$, this representation implements unitarity and analyticity and yields a controlled estimate of the $t$-channel contribution.
We employ two recent dispersive MO analyses~\cite{Lu:2020qeo,Deineka:2024mzt} that describe the existing $\gamma\gamma\to \pi\eta/K\bar{K}_{I_t=1}$ data. 
These minor differences stem from two sources: Ref.~\cite{Lu:2020qeo} employs a once-subtracted DR at the Adler zero ($t_0=t_A$) alongside a chiral $K$-matrix for $\pi\eta/K\bar{K}$, whereas Ref.~\cite{Deineka:2024mzt} relies on an unsubtracted dispersive representation with a data-driven $N/D$ approach.

The $S$-wave amplitudes $g_{+}^0(t)$ and $h_{+}^{0,1}(t)$ for $\pi \eta \rightarrow \bar{N} N$ and $\bar{K} K \rightarrow \bar{N} N$ obey the unitarity
\begin{align}\label{eq:unitarity_v3}
    \operatorname{Im} \mathbf{g}(t)=\mathbf{T}^*(t) \mathbf{\Sigma}(t) \mathbf{g}(t)\,, \quad \mathbf{g}(t)=\binom{g_{+}^0(t)}{\sqrt{2} h_{+}^{0,1}(t)}\,,
\end{align}
where $\mathbf{\Sigma}(t)=\operatorname{diag}\left(\sigma_{\pi\eta} \theta\left(t-t_{\pi\eta}\right), \sigma_{K\bar K} \theta\left(t-t_K\right)\right)$ with $\theta(x)$ the step function, and the convention for the $T$ matrix $\mathbf{T}$ follows Ref.~\cite{Lu:2020qeo}.
Following the Roy--Steiner equation analysis of nucleon form factors~\cite{Hoferichter:2012wf,Hoferichter:2016duk,Cao:2024zlf}, we write the twice-subtracted MO representation of $\mathbf{g}(t)$ as
\begin{align}\label{eq:g_DR}
    \mathbf{g}(t)={}& \boldsymbol{\Delta}(t)+\left(t-4m_N^2\right)\mathbf{\Omega}^1_0(t)\Bigg[\mathbf{P}_1(t)\nonumber\\
    &-\frac{t^2}{\pi} \int_{t_{\pi\eta}}^{\infty} \mathrm{d} t^{\prime} \frac{\operatorname{Im} \left[\mathbf{\Omega}^1_0\left(t^{\prime}\right)\right]^{-1} \mathbf{\Delta}\left(t^{\prime}\right)}{t^{\prime 2}\left(t^{\prime}-4m_N^2\right)\left(t^{\prime}-t\right)}\Bigg]\,.
\end{align}
At low energies, the inhomogeneity $\boldsymbol{\Delta}$ and the first-order subtraction polynomial $\mathbf{P}_1$ are matched to next-to-leading-order (NLO) SU(3) chiral perturbation theory (ChPT) amplitudes~\cite{Krause:1990xc,Oller:2006yh,Frink:2006hx,Ren:2012aj,Wu:2023uva,Wu:2024xwy}, supplemented by explicit decuplet baryons~\cite{Geng:2009hh}.
The matching between the dispersive representation and NLO ChPT is performed at the subtraction point $t=0$ by matching their expansions up to $\mathcal{O}(t)$. 
Although the decuplet resonance ($\Sigma^*$) is explicitly included as a degree of freedom, its low-energy polynomial components are systematically subtracted to avoid double counting with the chiral contact terms. 
A priori, partitioning between the left-hand cut nonpole pieces and the subtraction polynomial introduces an apparent scheme dependence~\cite{Garcia-Martin:2010kyn,Hoid:2026atr}. 
However, matching the full amplitude up to $\mathcal{O}(t)$ at $t=0$ ensures that the final dispersive result is manifestly scheme-independent and fulfills the low-energy chiral constraints; see Appendix~B of Ref.~\cite{Cao2026NP} for full technical details.

\begin{figure}[t]
    \centering
    \includegraphics[width=1\linewidth]{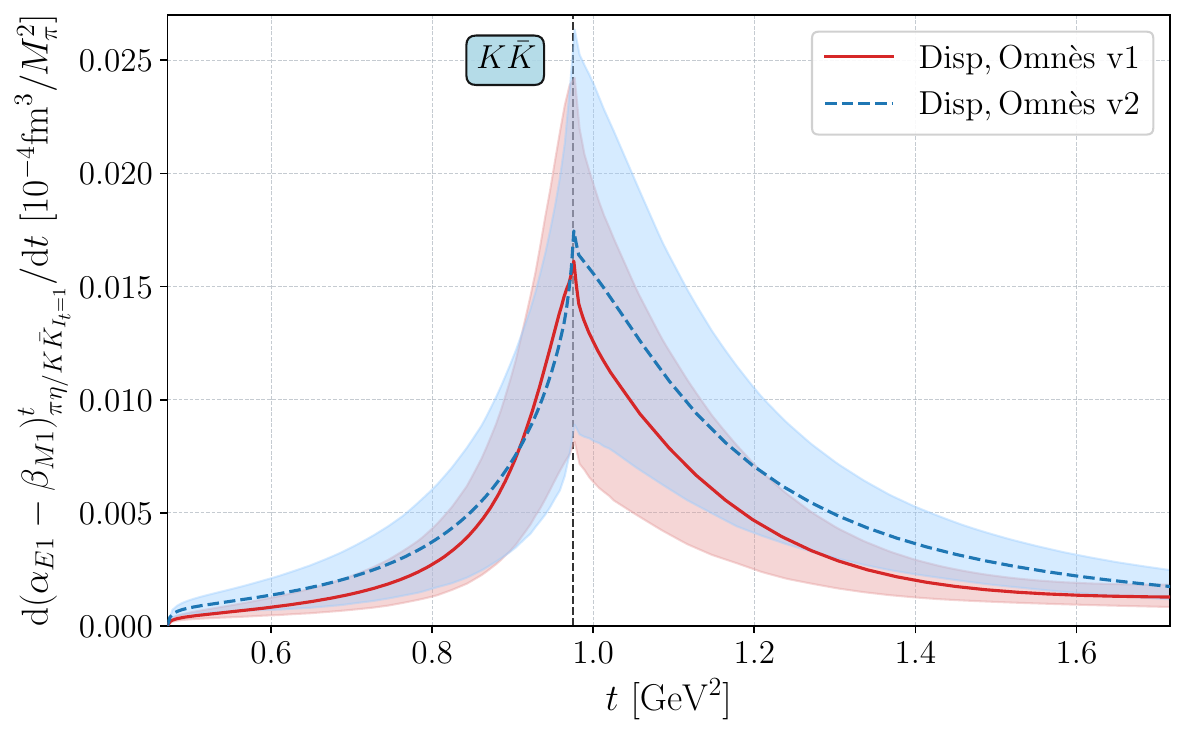}
    \caption{Integrand for the $t$-channel isovector contribution~\eqref{eq:SR_isovector_t}. The $S$-wave $I=1$ Omn\`es matrix is taken from Lu--Moussallam~\cite{Lu:2020qeo} (``Disp, Omn\`es v1'') and Deineka et al.~\cite{Deineka:2024mzt} (``Disp, Omn\`es v2''). We use fit~II of Ref.~\cite{Lu:2020qeo} and the solution corresponding to Eq.~(26) of Ref.~\cite{Deineka:2024mzt}. Shaded bands show only the uncertainty from the ChPT LECs, propagated by bootstrap.}
    \label{fig:t_a0}
\end{figure}

The two dispersive inputs agree well, as shown by the spectral function in Fig.~\ref{fig:t_a0}, despite their different $a_0(980)$ pole parameters~\cite{Lu:2020qeo,Deineka:2024mzt}.
Once unitarity and analyticity are imposed, the spectral function displays a pronounced cusp at the $K\bar{K}$ threshold.
We take the average of the two representations as the central value and estimate the $t$-channel uncertainty from three sources.
First, we use $(1.4~{\rm GeV})^2$ as the integral cutoff and estimate the associated uncertainty (``cut'') by varying it from $(1.3~{\rm GeV})^2$ to $(1.6~{\rm GeV})^2$ in Eq.~\eqref{eq:SR_isovector_t}; 
second, the small discrepancy between the central values of the two dispersive MO representations (``MO'') shown in Fig.~\ref{fig:t_a0} is treated as an uncertainty; 
third, the ChPT matching uncertainty (``ChPT'') is obtained by propagating the low-energy-constant (LEC) uncertainties~\cite{Ren:2012aj} with a bootstrap procedure, and the resulting bands are shown in Fig.~\ref{fig:t_a0}.
Adding the errors in quadrature gives $\left(\alpha_{E1}-\beta_{M1}\right)_{\pi\eta/K\bar{K}_{I_t=1}}^t=0.19(02)_\text{cut}(03)_\text{MO}(10)_\text{ChPT}=0.19(11)$.

Using Eq.~\eqref{eq:SR_isovector}, we find
\begin{align}\label{eq:a-b}
    \left(\alpha_{E1}-\beta_{M1}\right)^{p-n}=-0.82(33)\,,
\end{align}
where all errors are added in quadrature.
Comparing our result with the PDG average $\left(\alpha_{E1}-\beta_{M1}\right)^{p-n}=1.1(1.7)$~\cite{ParticleDataGroup:2026mpi} and the compilation in Ref.~\cite{Melendez:2020ikd}, $\left(\alpha_{E1}-\beta_{M1}\right)^{p-n}=-0.4(3.1)$, one finds that the uncertainty has been significantly reduced and a negative sign is determined at the $2.5\sigma$ level. 

\section{Implications for the proton--neutron mass difference}

The proton--neutron mass difference can be decomposed, to leading order in $e^2$, as~\cite{Gasser:1982ap,Gasser:2003hk}
\begin{align}
    m_p-m_n=\delta m_{\mathrm{QCD}}+\delta m_\mathrm{QED}+\mathcal{O}\left(e^4,e^2(m_d-m_u)\right)\,,
\end{align}
where $\delta m_{\mathrm{QCD}}$ is the QCD contribution driven by the light-quark mass difference $m_d-m_u$ and $\delta m_{\mathrm{QED}}$ is the electromagnetic contribution of order $e^2$.

The electromagnetic contribution is formally given by the Cottingham formula~\cite{Cottingham:1963zz}, which expresses it as an integral over the spin-averaged VVCS amplitude and a counterterm~\cite{Collins:1978hi,Hill:2016bjv,Gasser:2020hzn},
\begin{align}
    \delta m_\mathrm{QED}&=\lim_{\Lambda \rightarrow \infty}\left(m_\gamma^{\Lambda}+\Delta m^{\Lambda}\right)\,,\nonumber\\
    m_\gamma^\Lambda&=\frac{i e^2}{2 m_N} \int^\Lambda \frac{\md^4 q}{(2 \pi)^4} \frac{1}{q^2+i 0^+} T_\mu^\mu(p, q)\,,\nonumber\\
    \Delta m^{\Lambda}&=-\frac{3\alpha_{\mathrm{em}}}{8 \pi m_N}C \log \frac{\Lambda^2}{\mu^2}\,,
\end{align}
where $\Lambda$ is an ultraviolet regulator and the nucleon is at rest in the laboratory frame, $p^\mu =(m_N,\mathbf{0})$.
The counterterm $\Delta m^{\Lambda}$ absorbs the leading divergence arising from the renormalization of the mass difference. 
The small constant $C\simeq 6\times10^{-4}\,\mathrm{GeV}^2$~\cite{Collins:1978hi,Hill:2016bjv,Gasser:2020hzn} governs the asymptotic behavior of the VVCS amplitude.
The traditional renormalization of the Cottingham formula~\cite{Collins:1978hi} neglects Bjorken-scaling violations in the structure functions $F_{1,2}$ from short-distance singularities associated with spin-2 operators in the deep-inelastic region. 
A refined analysis~\cite{Gasser:2020hzn,Gasser:2020mzy} incorporates the perturbative QCD asymptotics of the VVCS amplitude, including such scaling violations, and allows $\delta m_{\mathrm{QED}}$ to be decomposed into three finite parts,
\begin{align}
    \delta m_{\mathrm{QED}}=m_{\mathrm{el}}+m_{F_{1,2}}+m_{\bar{S}}\,.
\end{align}
The term $m_{\mathrm{el}}$ is the well-determined elastic contribution from nucleon electromagnetic form factors.
Averaging over the parametrizations in Refs.~\cite{Kelly:2004hm,Punjabi:2015bba,Ye:2017gyb,Borah:2020gte} yields $m_{\mathrm{el}}=0.75(2)$~MeV.
The contribution from the structure functions $F_{1,2}$ is numerically suppressed, $m_{F_{1,2}}=-0.004(1)$~MeV~\cite{Gasser:2020hzn}. 
The dominant inelastic uncertainty resides in the term containing the subtraction function $\bar{S}\left(-Q^2\right)$, with $Q^2=-q^2$~\cite{Gasser:2020hzn},
\begin{align}
    m_{\bar{S}}=\frac{3\alpha_{\mathrm{em}}}{8 \pi m_N} \int_0^{\infty} \md Q^2 Q^2\left[\bar{S}\left(-Q^2\right)-\frac{C}{\left(\bar{\mu}^2+Q^2\right)^2}\right]\,,
\end{align}
with $\bar{\mu}=\mu \exp \left(-\frac{1}{2}\right)$, where $\mu=2$~GeV is the usual renormalization scale in the $\overline{\mathrm{MS}}$ scheme.

At $Q^2=0$, the subtraction function is related to $\left(\alpha_{E1}-\beta_{M1}\right)^{p-n}$ by a low-energy theorem~\cite{Bernabeu:1976jq,Pachucki:1999zza,Hill:2011wy,Gasser:2015dwa,Gasser:2020hzn},
\begin{align}\label{eq:Sbar0}
    \bar{S}(0)=-\frac{\kappa_p^2-\kappa_n^2}{4 m_N^2}+\frac{m_N}{2 \alpha_{\mathrm{em}}}\left(\alpha_{E1}-\beta_{M1}\right)^{p-n}\,,
\end{align}
where $\kappa_{p,n}$ are the proton and neutron anomalous magnetic moments.
With Eq.~\eqref{eq:a-b}, we find
\begin{align}\label{eq:Sbar}
    \bar{S}(0)=-0.56(28)~\mathrm{GeV}^{-2}\,.
\end{align}

\begin{figure}[t]
    \centering
    \includegraphics[width=1\linewidth]{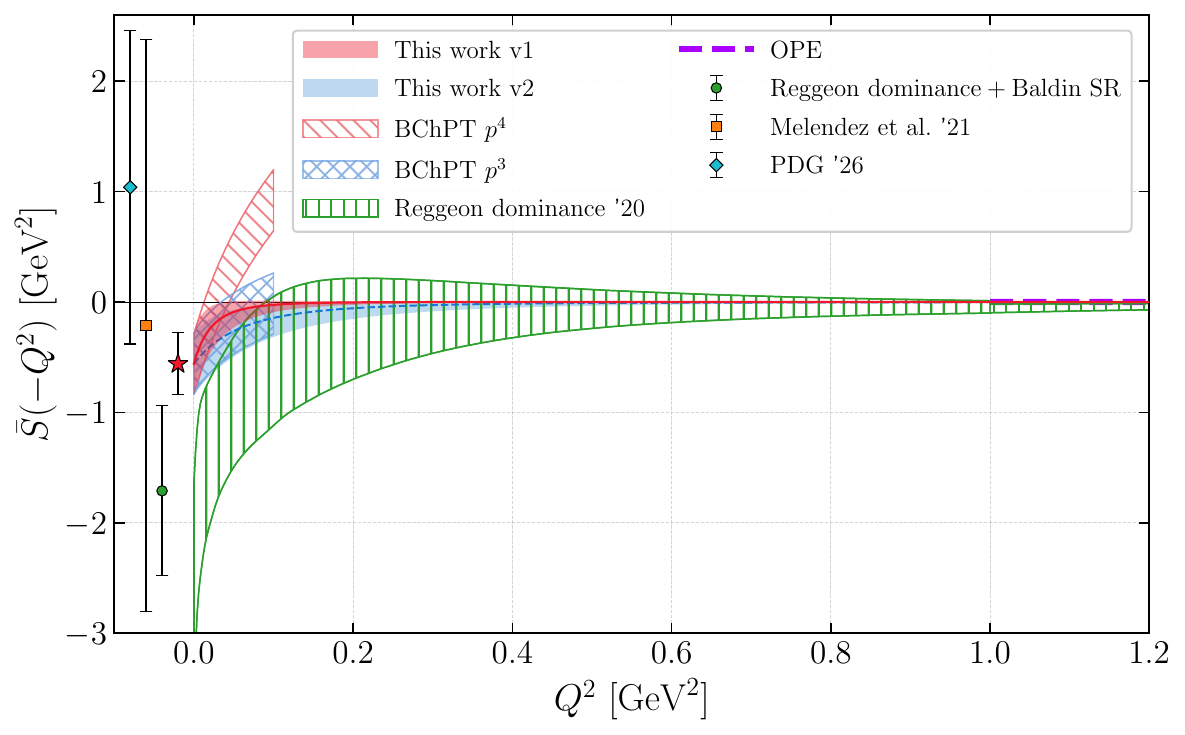}
    \caption{Subtraction function $\bar S(-Q^2)$ at low $Q^2$. The solid red (blue) band shows our result from Eqs.~\eqref{eq:Sbar} and~\eqref{eq:tripole} matched to $\mathcal{O}(p^4)$ [$\mathcal{O}(p^3)$] baryon ChPT (BChPT), while the red-hatched (blue-crossed) band denotes the corresponding standalone BChPT prediction~\cite{Lozano:2020qcg} anchored to $\bar S(0)$ from Eq.~\eqref{eq:Sbar}. The green vertical band represents the dispersive determination under Reggeon dominance~\cite{Gasser:2020hzn}; the purple dashed line shows the OPE asymptotic behavior~\cite{Collins:1978hi,Hill:2016bjv}. The error bar at $Q^2=0$ gives the total uncertainty of $\bar{S}(0)$. Also shown are the Reggeon-dominance prediction combined with the Baldin sum rule~\cite{Gasser:2020hzn}, the result of Melendez et al.~\cite{Melendez:2020ikd}, and the PDG average~\cite{ParticleDataGroup:2026mpi} at $Q^2=0$ via Eq.~\eqref{eq:Sbar0}, offset for clarity.}
    \label{fig:Sbar}
\end{figure}

Determining $m_{\bar{S}}$ requires the full $Q^2$ dependence of $\bar{S}(-Q^2)$, which could be inferred in a data-driven way if Reggeon dominance is assumed~\cite{Gasser:2020hzn}. 
Without making this assumption, Eq.~\eqref{eq:Sbar} fixes the crucial low-energy input, $\bar S(0)<0$. 
Recent lattice-QCD calculations leave little room for structure between the hadronic low-$Q^2$ region and the perturbative asymptotic domain: near the physical pion mass the isovector subtraction function has been computed for $Q^2\lesssim2$~GeV$^2$ and comes out negative and monotonic throughout~\cite{Fu:2024gxq}, while at the standard subtraction point the proton subtraction function already follows its perturbative asymptotics down to $Q^2\approx1$~GeV$^2$~\cite{Can:2025jzf}. 
Because $m_{\bar S}$ is dominated by $Q^2\lesssim2$~GeV$^2$ and the asymptotic tail contributes negligibly, $\bar S(0)<0$ then nearly dictates a negative subtraction contribution. 
As $m_{F_{1,2}}$ is negligible, $\delta m_{\rm QED}$ therefore lies below the elastic contribution $m_{\rm el}=0.75(2)$~MeV. 
To quantify this effect and facilitate comparison with previous Cottingham analyses, we use the tripole parametrization~\cite{Erben:2014hza} as a constrained benchmark for the interpolation between the chiral low-$Q^2$ region and the operator-product-expansion (OPE) regime, enforcing both $\bar{S}(0)$ and the asymptotic scaling $\bar{S}(-Q^2) \sim C/Q^4$~\cite{Collins:1978hi,Hill:2011wy,Gasser:2020hzn},
\begin{align}\label{eq:tripole}
    \bar{S}_{\mathrm{tripole}}\left(-Q^2\right)=\frac{\bar{S}(0)M_0^6+C Q^2}{\left(M_0^2+Q^2\right)^3}\,,
\end{align}
where $M_0=0.26^{+0.05}_{-0.05}$~GeV ($0.42^{+0.09}_{-0.11}$~GeV) is determined by matching Eq.~\eqref{eq:tripole} to the $\mathcal{O}(p^4)$ [$\mathcal{O}(p^3)$] ChPT result~\cite{Alarcon:2020wjg,Lozano:2020qcg} for $Q^2 \lesssim 0.1$~GeV.\footnote{The $\mathcal{O}(p^4)$ ChPT result depends on several poorly known LECs~\cite{Lozano:2020qcg}, which in practice can be constrained by $\bar S(0)$~\eqref{eq:Sbar}. 
Thus the ChPT input constrains the low-$Q^2$ slope of the subtraction function rather than $\bar S(0)$ itself.
An analogous treatment applies at $\mathcal{O}(p^3)$.}
The corresponding subtraction function is shown in Fig.~\ref{fig:Sbar}.
Evaluating $m_{\bar S}$ with Eq.~\eqref{eq:tripole}, we obtain $m_{\bar{S}}=0.00^{+0.00}_{-0.03}~\mathrm{MeV}$.

Combining this with $m_\mathrm{el}$ and $m_{F_{1,2}}$, we obtain
\begin{align}
    \delta m_\mathrm{QED}=0.75^{+0.02}_{-0.04}~\mathrm{MeV}\,.
\end{align}
This result is roughly consistent with Cottingham evaluations based on Reggeon dominance, $\delta m_\mathrm{QED}=0.58(16)$~MeV~\cite{Gasser:2020hzn,Gasser:2020mzy} within uncertainties. 
The uncertainty reduction follows because the improved low-$Q^2$ normalization fixed by the isovector polarizability tightens the constrained interpolation.
The difference from other Cottingham estimates~\cite{Walker-Loud:2012ift,Thomas:2014dxa,Erben:2014hza,Tomalak:2018dho} is largely traceable to the more precise isovector polarizability in Eq.~\eqref{eq:a-b}.
With the physical mass difference~\eqref{eq:CODATA}, the strong isospin-breaking contribution becomes
\begin{align}
    \delta m_\mathrm{QCD}=-2.04^{+0.04}_{-0.02}~\mathrm{MeV}\,.
\end{align}

\section{Conclusions}

In summary, we have shown that the nucleon isovector dipole polarizability combination $(\alpha_{E1}-\beta_{M1})^{p-n}$, whose sign remains unsettled, can be determined dispersively without invoking Reggeon dominance. 
The sum rule presented here relates it to $s$-channel photoabsorption cross sections and the $t$-channel amplitudes $\gamma\gamma \to \pi\eta/K\bar{K}_{I_t=1}$ and $\pi\eta/K\bar{K}_{I_t=1} \to N\bar{N}$. 
Using MO representations of the $t$-channel partial waves together with three empirical multipole data sets, we obtain $(\alpha_{E1}-\beta_{M1})^{p-n}=-0.82(33)\times10^{-4}\,\mathrm{fm}^3$, establishing a negative sign at the $2.5\sigma$ level with an uncertainty smaller than the previously known value by one order of magnitude.
The result provides an independent low-energy constraint, and hence a stringent consistency check, for Reggeon-dominance analyses of Compton amplitudes, complementary to current analyses based on data-driven functional extremal methods~\cite{Caprini:2016wvy,Caprini:2021btw}.

Through the Cottingham formula, $(\alpha_{E1}-\beta_{M1})^{p-n}$ fixes the low-energy normalization of the subtraction function that controls the dominant inelastic uncertainty in the electromagnetic proton--neutron mass difference, without the need to separately extract the neutron polarizabilities from nuclear Compton-scattering data. 
The most robust consequence is its sign: our result gives $\bar S(0)=-0.56(28)$~GeV$^{-2}$, implying a negative subtraction contribution  and therefore placing $\delta m_{\rm QED}$ below the well-known elastic contribution.
Within the class of constrained interpolations used in previous Cottingham studies~\cite{Walker-Loud:2012ift,Thomas:2014dxa,Erben:2014hza}, a tripole ansatz constrained by ChPT and the OPE then yields a precise value $\delta m_\mathrm{QED}=0.75^{+0.02}_{-0.04}$~MeV, with an uncertainty substantially smaller than that in the literature.

\bigskip

\begin{acknowledgments}
We are grateful to Bachir Moussallam for kindly sending the results of Ref.~\cite{Lu:2020qeo}, to Xiu-Lei Ren for sharing the correlation matrix of Ref.~\cite{Ren:2012aj}, and to Ulf-G.~Mei{\ss}ner for a careful reading of the manuscript and valuable comments.
In particular, XHC thanks Shi-Qing Kuang and Bing Wu for useful discussions.
This work is supported in part by National Natural Science Foundation of China under Grants No.~12125507, 12447101, 12322502, and 12335002, by the Chinese Academy of Sciences under Grant No.~YSBR-101, and by the Hunan Province Major Basic Research Projects with Grant No.~2026JC0006.

\end{acknowledgments}

\bibliography{refs}

\appendix

\setcounter{equation}{0}
\renewcommand{\theequation}{E\arabic{equation}}
\setcounter{figure}{0}
\renewcommand{\thefigure}{E\arabic{figure}}
\renewcommand{\theHfigure}{E\arabic{figure}}

\section{End Matter} \label{app:endmatter}

Here we provide details for the derivation of the isovector $t$-channel sum rule in Eq.~\eqref{eq:SR_isovector_t},
\begin{align}\label{eq:SR_isovector_t_app}
    &(\alpha_{E1}-\beta_{M1})^t_{\pi\eta/K \bar K_{I_t=1}}\nonumber\\
    ={}&\frac{8\alpha_\mathrm{em}}{\pi} \int_{t_{\pi\eta}}^{\infty}\frac{\mathrm{d} t}{t^2}\frac{1}{4 m_N^2-t}\sigma_{\pi\eta}(t) F_{\pi\eta,0\,0}(t) g_+^{0*}(t) \nonumber\\
    & +\frac{8\sqrt{2}\alpha_\mathrm{em}}{\pi} \int_{t_K}^{\infty}\frac{\mathrm{d} t}{t^2}\frac{1}{4 m_N^2-t}\sigma_{K\bar K}(t) F^{1}_{K,0\,0}(t) h_+^{0,1*}(t)\,.
\end{align}

To evaluate the $\pi\eta$ and $(K \bar K)_{I_t=1}$ intermediate-state contributions to the $t$-channel dispersion relation, we construct the partial-wave amplitudes for $\gamma \gamma \rightarrow M_1 M_2$ and $M_1 M_2 \rightarrow \bar{N} N$ using unitarity. 
We start by decomposing the $t$-channel helicity amplitude for $\gamma\gamma\to \bar{N} N$ into a partial-wave series,
\begin{align}
    T_{\lambda_{\bar{N}} \lambda_N, \lambda_\gamma \lambda_\gamma^{\prime}}^t(\nu, t)
    =&\sum_{J} \frac{2 J+1}{2}
    T_{\lambda_{\bar{N}} \lambda_N, \lambda_\gamma \lambda_\gamma^{\prime}}^{J(\gamma \gamma \rightarrow \bar{N}N)}(t)\nonumber\\
    &\times d_{\Lambda_\gamma \Lambda_N}^J\left(\theta_t\right)\,,
\end{align}
where $\Lambda_\gamma=\lambda_\gamma-\lambda_\gamma^{\prime}$, $\Lambda_N=\lambda_{\bar{N}}-\lambda_N$, and $\theta_t$ is the scattering angle in the $t$-channel center-of-mass frame, given by $\cos\theta_t={4 m_N \nu}/{\sqrt{t(t-4 m_N^2)}}$. 

By applying the Cutkosky cutting rule, the unitarity relation for the distinguishable intermediate states $M_1 M_2 \in \{\pi^0\eta, K^+K^-, K^0\bar{K}^0\}$ reads
\begin{align}\label{eq:unitarity_pieta/KK}
    &2 \operatorname{Im} T^{\gamma \gamma \rightarrow \bar{N}N} \nonumber\\
    =&\sum_{\{M_1 M_2\}}\frac{1}{(4 \pi)^2} \frac{p_{M_1 M_2}}{\sqrt{t}}
    \int \mathrm{d} \Omega_{M_1 M_2}\nonumber\\
    &\times\left[T^{\gamma \gamma \rightarrow M_1 M_2}\right]
    \left[T^{M_1 M_2 \rightarrow \bar{N}N}\right]^*\,,
\end{align}
where $p_{M_1 M_2} = \sqrt{\lambda(t,M_1^2,M_2^2)}/(2\sqrt{t})$ is the center-of-mass three-momentum of the $M_1 M_2$ system, and the invariant phase-space factor is defined as $\sigma_{M_1 M_2}(t) \equiv 2p_{M_1 M_2}/\sqrt{t}$. 
Up to azimuthal phases absorbed into the Wigner rotation matrices, the partial-wave expansions for the subprocesses $\gamma\gamma \rightarrow M_1 M_2$ and $M_1 M_2 \rightarrow \bar{N} N$ are
\begin{align}
    &T_{\Lambda_\gamma}^{\gamma \gamma \rightarrow M_1 M_2}\left(t, \theta_{M_1 M_2}\right) \nonumber\\
    =& \sum_{J~\mathrm{even}} \frac{2 J+1}{2}
    T_{\Lambda_\gamma}^{J\left(\gamma \gamma \rightarrow M_1 M_2\right)}(t)
    \sqrt{\frac{\left(J-\Lambda_\gamma\right)!}{\left(J+\Lambda_\gamma\right)!}}\nonumber\\
    &\times P_J^{\Lambda_\gamma}\left(\cos \theta_{M_1 M_2}\right)\,, \label{eq:PW_gg->pieta/KK} \\
    &T_{\Lambda_N}^{M_1 M_2 \rightarrow \bar{N}N}(t, \Theta) \nonumber\\
    =& \sum_J \frac{2 J+1}{2}
    T_{\Lambda_N}^{J(M_1 M_2 \rightarrow \bar{N}N)}(t)
    \sqrt{\frac{\left(J-\Lambda_N\right)!}{\left(J+\Lambda_N\right)!}}\nonumber\\
    &\times P_J^{\Lambda_N}(\cos \Theta)\,. \label{eq:PW_pieta/KK->NN}
\end{align}
Combining Eqs.~\eqref{eq:PW_gg->pieta/KK} and~\eqref{eq:PW_pieta/KK->NN} and integrating over the solid angle in Eq.~\eqref{eq:unitarity_pieta/KK} via the orthogonality of rotation matrices, we deduce the imaginary parts of the $t$-channel partial waves:
\begin{align}\label{eq:unitarity_pw_pieta/KK}
    &2 \operatorname{Im} T_{\lambda_{\bar{N}} \lambda_N, \lambda_\gamma \lambda_\gamma^{\prime}}^{J\left(\gamma \gamma \rightarrow \bar{N}N\right)}(t) \nonumber\\
    =&\sum_{\{M_1 M_2\}}\frac{1}{8 \pi} \frac{p_{M_1 M_2}}{\sqrt{t}}
    \left[T_{\Lambda_\gamma}^{J\left(\gamma \gamma \rightarrow M_1 M_2\right)}(t)\right]\nonumber\\
    &\times\left[T_{\Lambda_N}^{J(M_1 M_2 \rightarrow \bar{N}N)}(t)\right]^*\,.
\end{align}

These partial waves are related to the standard reduced amplitudes as follows. 
For the $\Lambda_N=0$ proton--antiproton ($\bar{p}p$) final state, the strong partial-wave amplitudes are related to the Frazer--Fulco-type amplitudes $g_{+}^J(t)$ (defined for the physical $\pi^0\eta\to\bar{p}p$ channel) and the isospin-decomposed kaon amplitudes $h_{+}^{J,I}(t)$ ($I=0,1$) by
\begin{align}\label{eq:FF_amp}
    T_{\Lambda_N=0}^{J(\pi^0 \eta \rightarrow \bar{p}p)}(t) &= \frac{16 \pi}{p_N} \left(p_N p_{\pi\eta}\right)^J g_{+}^J(t), \nonumber\\
    T_{\Lambda_N=0}^{J(K^+ K^- \rightarrow \bar{p}p)}(t) &= \frac{16 \pi}{p_N} \left(p_N p_{K\bar K}\right)^J \left[h_{+}^{J,0}(t) - h_{+}^{J,1}(t)\right], \nonumber\\
    T_{\Lambda_N=0}^{J(K^0 \bar{K}^0 \rightarrow \bar{p}p)}(t) &= \frac{16 \pi}{p_N} \left(p_N p_{K\bar K}\right)^J \left[h_{+}^{J,0}(t) + h_{+}^{J,1}(t)\right]\,,
\end{align}
where $p_N=\sqrt{t / 4-m_N^2}$ is the nucleon center-of-mass momentum. The photonic partial-wave amplitudes are normalized as
\begin{align}\label{eq:gg_amp_v1}
    T_{\Lambda_\gamma}^{J(\gamma \gamma \rightarrow M_1 M_2)}(t)=\frac{4 e^2}{\sqrt{2 J+1}} F_{M_1 M_2, J \Lambda_\gamma}(t)\,.
\end{align}
Following the kaon charge conventions of Ref.~\cite{Dai:2014zta}, the charged ($\gamma\gamma\to K^+ K^-$) and neutral ($\gamma\gamma\to K^0 \bar{K}^0$) two-photon amplitudes decompose into isoscalar ($F_K^0$) and isovector ($F_K^1$) components as
\begin{align}\label{eq:gg_K_isospin}
    F_{K, J\Lambda_\gamma}^C &= -\frac{F_{K, J\Lambda_\gamma}^0 + F_{K, J\Lambda_\gamma}^1}{\sqrt{2}}\,, \nonumber\\
    F_{K, J\Lambda_\gamma}^N &= -\frac{F_{K, J\Lambda_\gamma}^0 - F_{K, J\Lambda_\gamma}^1}{\sqrt{2}}\,.
\end{align}
Summing over the charged and neutral kaon intermediate states in Eq.~\eqref{eq:unitarity_pw_pieta/KK} therefore yields the combination
\begin{align}\label{eq:gg_amp_v2}
    &F_{K, J\Lambda_\gamma}^C \left[h_+^{J,0} - h_+^{J,1}\right]^* + F_{K, J\Lambda_\gamma}^N \left[h_+^{J,0} + h_+^{J,1}\right]^* \nonumber\\
    =& -\sqrt{2} F_{K, J\Lambda_\gamma}^0 h_+^{J,0*} + \sqrt{2} F_{K, J\Lambda_\gamma}^1 h_+^{J,1*}\,,
\end{align}
which shows that the isoscalar ($I=0$) and isovector ($I=1$) kaon contributions enter with opposite signs and a characteristic relative weight of $\sqrt{2}$.

In the coupled-channel unitarity formulation for the isovector $S$-wave, this relative $\sqrt{2}$ factor is naturally accommodated by defining the strong amplitude vector as $\mathbf{g}(t) = \big(g_+^0(t),\, \sqrt{2}h_+^{0,1}(t)\big)^T$.
With this normalization, the $\pi\eta$ and $K\bar{K}_{I_t=1}$ channels share a standard invariant scattering matrix $\mathbf{T}$ satisfying the matrix unitarity relation $\operatorname{Im} \mathbf{g}(t) = \mathbf{T}^*(t)\mathbf{\Sigma}(t) \mathbf{g}(t)$.

In the backward scattering limit ($\nu^2=tp_N^2/(4m_N^2)$, i.e., $\cos\theta_t = \pm 1$), the invariant L'vov amplitude $\tilde{A}_1 = A_1 - \frac{t}{4m_N^2}A_5$ isolates the helicity-conserving ($\Lambda_\gamma=\Lambda_N=0$) amplitudes via 
\begin{equation}
    \tilde{A}_1(\nu_t, t) = \frac{T^t_{\frac{1}{2}\,\frac{1}{2},1\,1}+T^t_{\frac{1}{2}\,\frac{1}{2},-1\,-1}}{t\sqrt{t-4m_N^2}}\,.
\end{equation}
Using parity conservation ($T^t_{\frac{1}{2}\,\frac{1}{2},1\,1} = T^t_{\frac{1}{2}\,\frac{1}{2},-1\,-1}$ for even $J$) and analytic continuation into the pseudophysical region $t < 4m_N^2$ where $p_N^2 = -(4m_N^2 - t)/4$, we substitute Eqs.~\eqref{eq:FF_amp}--\eqref{eq:gg_amp_v2} into Eq.~\eqref{eq:unitarity_pw_pieta/KK} for the $S$-wave ($J=0, I=1$) component to obtain
\begin{align}
    &\operatorname{Im}_t \tilde{A}_1^{(I=1, J=0)}(\nu_t, t)\nonumber\\
    =& -\frac{16\pi\alpha_\mathrm{em}}{t(4m_N^2 - t)}\Big[ \sigma_{\pi\eta}(t) F_{\pi\eta,0\,0}(t) g_+^{0*}(t) \nonumber\\
    &\qquad\qquad\qquad+ \sqrt{2}\sigma_{K\bar{K}}(t) F_{K,0\,0}^1(t) h_+^{0,1*}(t) \Big]\,,
\end{align}
with $e^2 = 4\pi\alpha_\mathrm{em}$ and step functions $\theta(t-t_{\pi\eta})$, $\theta(t-t_K)$ implicit in the phase spaces. Substituting this result into the backward $t$-channel dispersion integral
\begin{align}\label{eq:a-b_int}
    (\alpha_{E1}-\beta_{M1})^t=-\frac{1}{2 \pi^2} \int_{t_{\text{th}}}^{\infty}\frac{\mathrm{d} t}{t}~ \operatorname{Im}_t \tilde{A}_1\left(\nu_t, t\right)
\end{align}
directly yields the isovector sum rule in Eq.~\eqref{eq:SR_isovector_t_app}.

\end{document}